# Rethinking Reproducibility in the Classical (HPC)-Quantum Era: Toward Workflow-Centered Science

Anna Vrtiak[1], Duuk Baten[1], Ariana Torres-Knoop[1, *]

[1]SURF B.V, Moreelsepark 48, Utrecht, The Netherlands

**Abstract**

Scientific knowledge increasingly depends on complex computational processes where both hardware and software layers can influence research outcomes. As computational complexity grows, classical-quantum integration provides a lens for examining how the scientific method adapts, particularly regarding a foundational principle of scientific validation - reproducibility. Building upon previous warnings of an ongoing reproducibility crisis in the computational context, this paper examines challenges across classical (HPC) and quantum computing. Despite its deterministic nature, HPC faces reproducibility threats from hardware dependencies, documentation inadequacies, disincentivizing research culture and infrastructure variation. Quantum computing, at low technological maturity, amplifies some challenges, while creating new ones through probabilistic outputs, hardware-specific noise, and tight software-hardware coupling. Classical-quantum integration reveals a telling pattern, where current reproducibility frameworks prove inadequate, as infrastructure blends with the results. Quantum integration serves as a catalyst exposing methodological limitations across the computational domain. We propose a workflow-centered path forward, pointing to the value of gradual cultural shift toward workflow-centered scientific practice. By developing meta-workflows that document both process abstractions and implementation contexts, we create a more robust foundation for scientific knowledge that acknowledges complexity without sacrificing rigor. The path forward involves embracing this evolution in understanding scientific knowledge rather than resisting it.

## 1. Introduction

As quantum computers begin integrating with high-performance computing (HPC) as accelerators [1], they introduce additional complexity that strains established methods for ensuring scientific rigor. With research across fields - from astrophysics and genomics to climate science and materials design [2] - now increasingly dependent on computational infrastructure, maintaining scientific credibility requires new approaches for validation that can accommodate this quantum-HPC integration as well as the broader evolution of computational heterogeneity and complexity.

HPC has historically powered some of the most demanding scientific applications, enabling breakthroughs in aerospace, life sciences, finance, and energy [3]. While the deterministic nature of classical computers suggests that computational research should achieve high reproducibility standards, scientists often cannot reproduce computational findings due to complexities in software packaging, installation, and execution, combined with limitations in how these processes are documented [4].

Among the scientific principles affected by the increasing complexity in computation, reproducibility stands out as particularly vulnerable [5]. Reproducibility – the ability of independent researchers to obtain consistent results using the same methods and experimental setup [6] – has long been upheld as a cornerstone of scientific credibility [7]. In the computational context, reproducibility encompasses diverse scenarios: one attempt to reproduce might involve running the same version of the software on a new version of an operating system, while another attempt to reproduce might involve writing a new piece of software that implements the same algorithm [5].

What constitutes "the same setup" in computational contexts is far from straightforward, as seemingly equivalent approaches can yield different results due to underlying environmental differences. This problem is particularly evident in HPC environments, where software may be upgraded, removed, or recompiled without users' knowledge, making it difficult to reproduce results even on the same system over time [8]. Additionally, HPC performance optimization often requires commands tailored to specific hardware configurations, making infrastructure independence practically challenging [5].

These challenges highlight the rising role of computational workflows – structured sequences of steps composed of control and data flow statements and rules which perform the analytics required to achieve the intended experimentation [9]. Reflecting the broader shift toward computational methods across scientific domains, workflows have become integral to the scientific method. What were once simple execution scripts have transformed into complex mechanisms for embodying scientific processes themselves [10]. Their importance intensifies when computational infrastructure becomes inseparable from the experiment itself, a dynamic particularly evident in quantum computing.

---

* ariana.torres@surf.nl

While quantum computing is emerging as a powerful addition to the classical computational infrastructure, it not only amplifies these existing challenges but also generates novel challenges of its own. Unlike their classical counterparts, quantum computers produce inherently probabilistic outputs rather than deterministic results, meaning that identical quantum programs can yield different statistical distributions across runs [11]. At their current low technological maturity (TRL) [12], they also exhibit hardware-specific noise profiles, with distinct physical architectures that can produce varying results even for the same algorithm [13].

The upcoming integration, resulting in new classical-quantum infrastructures, thereby raises new questions for scientific methodology e.g:

- *What does reproducibility mean when computational outcomes are probabilistic by nature?*
- *How can we document workflows that involve highly specialized hardware with device-specific, fluctuating behavior?*
- *How do we ensure scientific rigor when computational infrastructure becomes an integral part of the experimental apparatus?*

This paper addresses these questions by examining reproducibility challenges across classical and quantum computing and explores their integration. We propose workflow-centered science as a path forward, with meta-workflows as higher-level abstract representations of research logic that transcend specific infrastructure implementations [9]. We argue that this approach is necessary not only for the classical-quantum integration but for strengthening scientific rigor across all computational research in an era of ever-increasing technological complexity.

## 1.1 Reproducibility: Crisis, Definitions, and the Computational Context

Before examining how classical-quantum integration challenges scientific validation, we must establish what reproducibility means in (computational) research and why ensuring it has become increasingly difficult. The inability to reproduce scientific findings has emerged as a critical concern, with surveys revealing that over 70% of researchers have failed to reproduce another scientist's experiments, and more than 50% cannot reproduce even their own previous work [2]. What has come to be known as the *reproducibility crisis* extends beyond methodological concerns, affecting policy decisions, research funding, and public confidence in scientific institutions [14]. In response, the scientific community has developed frameworks to strengthen research practices, with Open Science initiatives promoting transparency in research processes, data sharing, and electronic documentation [15, 16]. Notably, the community-driven FAIR principles – making research Findable, Accessible, Interoperable, and Reusable – providing guidelines for improving reproducibility, have become particularly popular in scientific discourse.

As we dive into what is at stake when ensuring reproducibility in the world of computation, the first concern begins already by terminological confusion complicating these efforts. Distinct scientific disciplines use "reproducibility" and "replicability" inconsistently, and often contradictorily [17]. Following recent standardization [6], this paper adopts the ACM definitions:

**Reproducibility**: Different team, same experimental setup. Independent researchers using identical methods and conditions should obtain consistent results.

**Replicability**: Different team, different experimental setup. Independent researchers using different methods or data should reach the same scientific conclusions.

In the context of computational research, reproducibility means that researchers should be able to obtain consistent results with the same input, computational steps, code, and conditions [18]. Yet, this ideal breaks down due to layers of complexity inherent to the computational context, including differing software versions, hardware dependencies, operating system variability, random number generation, and floating-point precision issues, among many others [7].

As a result, in computation, reproducibility is both a technical and epistemic challenge. Research results depend on layers of technology that are often invisible to researchers but directly affect outcomes, calling for examining reproducibility's foundations. The next paragraphs establish the theoretical framework (robustness) and practical dimensions (three axes) needed to analyze how reproducibility works across classical and quantum computing.

## 1.2 Foundations of Reproducible Science

To understand reproducibility meaningfully, robustness provides a conceptual foundation. As outlined by Karaca [19], coming from the philosophy of science, *robustness* tells us why some scientific findings are considered more reliable than others by looking at both the stability and reliability of experimental procedures and results under varying conditions. It can be divided into:

*Procedure Robustness (PR):* The capacity of an experimental procedure to maintain its intended function despite variations in inputs or conditions. A robust procedure produces reliable outputs even when faced with minor fluctuations in experimental conditions, measured entities, or implementation details.

*Results Robustness (RR):* Obtaining consistent or convergent results through different and independent means of detection or methodological approaches. If RR holds, we gain confidence that those results reflect the underlying reality rather than artifacts of a particular methodology.

Importantly, PR is a precondition for RR, together forming a precondition for meaningful reproducibility. If procedures lack robustness, different researchers may get inconsistent results, making it impossible to determine whether original findings are valid. The complexity of modern computing threatens both PR and RR. PR requires workflows that tolerate environmental variations, such as producing stable outputs across different operating systems, software versions, or numerical libraries. RR strengthens confidence when different models, algorithms, or implementations converge on the same conclusions, validating that findings reflect genuine phenomena rather than methodological artifacts.

What follows from the *robustness* analysis is to ask: *How can we structure our computational research to maximize stability and consistency?* To bridge the gap between theory and practice, in the next subsection, we evaluate reproducibility along three interconnected axes that together provide a comprehensive evaluation tool.

## 1.3 Computational Reproducibility: Dimensions and Infrastructure Awareness

To move from conceptual understanding to practical implementation of reproducibility in computational science, it is useful to adapt Chirigati et al. 's three complementary axes of reproducibility [20]:

*Transparency* addresses the openness of research materials. As transparency increases, researchers share complete datasets, executable code, and eventually entire computational environments, enabling other researchers to examine and validate the processes that transformed raw data into published findings. Computational transparency emerges when researchers can trace the path from input to output by making both data and executable scripts and programs available

*Portability* examines how widely research can be reproduced across different computing environments. Research with low portability requires the original system, while highly portable research runs consistently across different operating systems and hardware architectures. Designing for portability requires careful consideration of these environmental factors.

*Coverage* evaluates how much of the research pipeline is captured in reproducible form. Partial coverage allows reproduction of isolated steps but leaves gaps in the workflow. Full coverage captures every step from raw data acquisition through analysis to final reporting, creating an unbroken chain of reproducible processes. This complete workflow documentation ensures that every transformation from initial data to reported conclusion can be examined and potentially extended by other researchers.

Taken together, these three axes interact to determine the overall reproducibility of computational research. The most robustly reproducible research maximizes all three dimensions, yielding work that is transparent, widely portable, and thoroughly documented across the entire research pipeline.

## 1.4 Infrastructure Variation and Functional Conservation

The growing complexity and heterogeneity of computational environments, coupled with the rapid evolution of software ecosystems and operating systems, have significantly intensify the challenge of ensuring reproducibility in modern scientific computing. The proliferation of software versions, dependencies, and orchestration layers creates an increasingly intricate and unstable execution landscape. These problems are further amplified in advanced computational environments such as HPC centers, where users do not have full control A previous study [20] found that nearly 80% of sampled workflows in HPC context failed to be reproduced, with approximately 12% of these failures occurring due to missing execution environment specifications. Without proper documentation of the infrastructure layer (hardware + software environment including orchestration), scientists attempting to reproduce an experiment are forced to infer the original configuration.

A central conceptual challenge concerns defining acceptable variation under the "same setup" condition of computational reproducibility. With the rapid technology pace, maintaining identical HPC environments indefinitely is neither practical nor desirable. Instead, the objective should be toward *logical conservation* [21]: preserve the functional capabilities and configurations necessary for reproducing research, rather than physically preserving exact systems. Importantly, not all infrastructure variations affect scientific reproducibility equally. Establishing benchmarks for what constitutes an acceptable level of infrastructure variation requires distinguishing between elements that may affect scientific outcomes and those that merely impact performance or operational convenience. Such benchmarks would then create a point against which reproducibility efforts can be evaluated, ensuring that critical execution characteristics are preserved while allowing non-essential elements to evolve alongside technological progress. Specific examples and proposed criteria are explored in the following section based on the analysis of common sources of reproducibility loss.

## 2. Reproducibility in Classical Computing

To effectively address reproducibility in computational environments, we need to understand the factors that undermine it. *Table 1* presents an overview of common reproducibility challenges in classical computing, drawing from literature review [7] and expert focus groups at SURF, the Dutch IT cooperation for research and education [22]. The outlined challenges span three categories: 1) technical issues arising from the computational stack itself, 2) human factors related to research practices and documentation and 3) organizational structures that shape incentive systems and cultural norms.

*Table 1.* Reproducibility Challenges in Classical Computing

| Challenges to Reproducibility | Examples | Reproducibility | Infrastructure Variation (Same set-up condition) |
|---|---|---|---|
| **Technical Challenges** | | | |
| Software Environment | Software backwards compatibility | ● | ● |
| | Convoluted software layers | ● | ● |
| | Snapshot/Versioning of environments | ● | ● |
| | Software libraries outside user's control | ● | ● |
| Workflow | Workflow change logs/Lack of transparency | ● | ● |
| | Different initial/boundary conditions | ● | ● |
| | Unclear separation between software & workflow | ● | ● |
| Silent Errors | | ○ | ● |
| Machine Learning | | | ● |
| Hardware Heterogeneity | Access to different hardware types | ● | ● |
| | Numerical drift due to infrastructure changes | | ● |
| PRNGs | | ○ | ● |
| Parallel Computing | Numerical precision | ○ | ● |
| Optimization | Numerical instability | ○ | ● |
| | Compilers | ○ | ● |
| | Optimization not visible to user | ○ | ● |
| Rapid Development | | | ● |
| Data Management | Large datasets disappearing | ● | ● |
| Quantum Computing | *(addressed in Section 2)* | ● | ● |
| SE + Documentation | Modifications in SW not tracked | ● | ● |
| **Human Factors** | | | |
| Software Engineering as a Black box | Lack of expertise in reproducible scientific computing | ● | |
| | Lack of training in computation sciences | ● | |
| Documentation | Team member leaving | ● | |
| | Cross-cultural Formal vs. Informal Documentation Practices | ● | ○ |
| | No clear ownership for documenting infrastructure changes | ● | ● |
| Statistics (P-value hacking, HARKing) | Quick-to-solution instincts | ● | |
| **Org./Systemic issues** | | | |
| Scientific Culture | Lack of incentives / no reward for reproducibility | ● | |
| | Novelty bias: Cannot publish work in high-tier journals | ● | |
| | Lack of systemic awareness on the issue | ● | |
| | Reluctance to share due to critique | ● | |
| | Culture never developed due to previously stable infrastructure | ● | |
| Degree of Openness of Science | Lack of standards/guidelines for documentation | ● | |
| | Infrastructure/software developed by vendors without researcher say | ● | |

Legend: ● = Primary threat    ○ = Secondary Threat    (Blank) = No significant threat

*Table 1* further distinguishes between two types of impact. Some challenges directly threaten reproducibility itself, while others can create infrastructure variation concerns, affecting the technical definition of "same setup" – what delta of environment change remains acceptable without invalidating scientific findings, from which the boundaries of acceptable variation in computational environments are formed. The table categorizes challenges by their primary impact, based on expert assessment: primary threats (●), secondary threats (○), or no significant threat (blank).

All things considered, reproducibility challenges rarely have a single cause. For instance, a software version change (technical) may go undocumented (human) because no institutional incentive exists to maintain detailed records (organizational). Examining this interconnectivity closer reveals why technical solutions alone cannot resolve reproducibility challenges, where the problem requires coordinated attention across all three domains. This naturally also raises questions about responsibility, spanning across the technical, human, and organizational domains.

Different aspects of reproducibility fall under different stakeholders - in this paper, we identify three groups, with responsibility distributed across: (1) *Researchers*,

(2) *Infrastructure Providers* (such as HPC centers), and (3) *Scientific Institutions and Communities.*

According to the expert focus groups, *Researchers* must thoroughly document their scientific methodology: software dependencies, parameter selections, data preprocessing, and analytical decisions. Computational methods demand the same rigor as physical experiments, with reproducibility built into research design from the start, instead of being retrofitted later. *Infrastructure providers* should document computing environments, including system architectures, hardware specifications, and software configurations. They can support reproducibility efforts by developing tools that automatically capture environmental information, creating standardized environments, and providing training on best practices. *Scientific institutions and communities* actively shape reproducibility norms through incentives, policies, and education. In this light, they should establish guidelines, create recognition mechanisms for reproducible work, allocate resources, and incorporate reproducibility into training programs.

In reality, boundaries blur, for example when researchers bring custom software to HPC systems. The distinction between user and infrastructure responsibilities becomes unclear, requiring flexible accommodation of varying responsibility distributions depending on specific use cases. Effective reproducibility demands collaboration across all stakeholders, acknowledging that modern computational science exceeds any single researcher's or institution's expertise

## 3. Reproducibility in Quantum Computing

In classical high-performance computing, many (technical) reproducibility challenges largely arise from infrastructural complexity: heterogeneous hardware, evolving software stacks, dependency management, compiler behavior, and orchestration layers. Although difficult to manage, these sources of variation are fundamentally deterministic. Given sufficient documentation and specification of the execution environment, identical methods and conditions (within the acceptable variation), should yield consistent results. In quantum computers, variability exists at a more fundamental level.

Quantum systems operate on the principles of superposition, interference and entanglement. They encode and process information in a different manner offering potential advantages over classical computing, the so-called *quantum advantage* [24]. However, the same quantum features that bring promise to quantum computers, introduce engineering and reproducibility challenges. Quantum systems must remain isolated to maintain coherence, yet must also be controllable and readable, requirements that are often in tension with one another [11]. Even minor environmental disturbances can impact results, and running the same program on two devices with similar specifications may yield different outcomes.

### 3.1 Technical challenges

**Low Technology Readiness Level**

Quantum computing currently remains at approximately Technology Readiness Level (TRL) 3 [12], transitioning from proof-of-concept demonstrations to early prototypes [25]. Operational standards, benchmarking conventions, and best practices are still evolving. We are currently in what physicist John Preskill termed the *NISQ era: Noisy Intermediate-Scale Quantum* [11]. NISQ devices have approximately fifty to a few hundred noisy qubits. These devices are shaped by computational errors from hardware noise, imperfect gate operations, and environmental interference [13], directly limiting circuit depth and result reliability. While the long-term objective is the realization of fault-tolerant, scalable quantum computers capable of suppressing such variability through error correction, this goal remains technologically distant. In the meantime, reproducibility in near-term quantum systems must contend with hardware instability.

**Hardware Diversity**

Unlike classical computing, which benefits from relatively standardized processor architectures, quantum hardware is highly diverse, encompassing a range of architectures such as superconducting circuits, trapped ions, photonic qubits, neutral atoms, and spin-based systems [26]. Each architecture has distinct physical and engineering properties and constraints. Standardization across platforms is minimal, and there is no universal hardware baseline. Even devices based on the same architecture may differ significantly in practice.

**Noise as structural feature**

Quantum systems are affected by multiple interacting noise sources, whose relative contributions are often difficult to isolate [27]. A central challenge lies in the coherence–controllability tradeoff: qubits must remain sufficiently isolated to preserve quantum states while simultaneously interacting with control systems to perform computation. Because these noise sources drift over time and vary across devices, reproducibility cannot be guaranteed by merely replicating code. The physical implementation matters. Some major noise sources include:

- **Decoherence ($T_1$ and $T_2$ limits):** The coherence properties of a qubit are characterized by the decoherence times $T_1$ (relaxation time) and $T_2$ (dephasing time). Once these timescales are exceeded, state fidelity deteriorates, introducing stochastic errors that limit computational reliability.
- **Imperfect gate operations:** Imperfect quantum gate implementations introduce additional noise. Even small systematic angular deviations can alter output distributions and are central parameters in quantitative assessments of circuit reproducibility [13]. These errors are particularly problematic for reproducibility because they often produce coherent, systematic deviations rather than purely random noise. Since these control errors depend on calibration procedures that vary across time and between labs, two researchers executing the identical circuit description could obtain different results because the physical implementation of each gate operation varies across devices and calibration cycles.
- **Readout errors:** Measurement errors (e.g. mislabeling $|0\rangle$ state as $|1\rangle$) vary across qubits and over time, directly affecting the statistical outputs that constitute experimental results in quantum computing.
- **Crosstalk:** Unintended interactions between neighboring qubits during operations introduce layout-dependent errors. As device topology differs across platforms, crosstalk further complicates cross-hardware reproducibility.

The illustrated hardware challenges to reproducibility extend beyond the physical layer. At this TRL, quantum software remains tightly coupled to specific hardware implementations, translating noise sources and device-specific behaviors into the software domain and compounding reproducibility losses. High-level quantum programming languages (such as Qiskit [28] or Cirq [29]) offer abstraction, but all code eventually gets compiled through a stack of transpilers that optimize for specific hardware. The same circuit written in Qiskit could be compiled differently depending on the hardware architecture (even with identical parameters), as outlined in the previous section. By the same token, two researchers using identical high-level code on different quantum processors - or even the same processor at different times - could receive entirely different compiled circuits with different error profiles and results. Thus, without detailed knowledge of the entire compilation stack and hardware configuration, reproducing another researcher's results seems hardly possible.

This resembles lessons learned from classical HPC, where dependency versions and compiler configurations influence reproducibility. However, in quantum computing, compilation is not merely a performance optimization step; it directly shapes the physical realization of computation. When researchers document experiments that use quantum computing, their focus should not lie only on the high-level algorithm description, but also hardware-specific compilation details and software configurations should not be omitted. In this light, reproducibility efforts require comprehensive metadata about the entire software stack. However, even when researchers document their experimental setup as precisely as possible, cross-platform reproducibility remains limited. This is tied to the lack of standardized hardware usage, with each type requiring different software approaches, creating more silos rather than far-reaching efforts in reproducible research.

### 3.2 Human & Cultural Factors

Overall, the technical aspects of quantum reproducibility discussed thus far – hardware noise, software specificity, and lack of standardization efforts – represent only part of the reproducibility considerations. Beyond technical aspects, human and cultural factors likewise impact reproducibility in quantum computing research, as is the case in classical computing.

Quantum computing research is not deployed in a vacuum and is inherently shaped by the organizational and systemic factors shaping scientific practices. These, in turn, shape how researchers interact with quantum technologies and can determine whether reproducibility becomes a priority or an afterthought. The path toward reproducible quantum computing research requires recognizing the non-technical dimension, and an examination of whether the human and cultural factors amplify or mitigate reproducibility efforts.

Quantum computing is inherently interdisciplinary, spanning physics, computer science, and application domains. Similar to classical computing, researchers may treat computational tools as black boxes. However, quantum systems intensify this opacity:

domain scientists may lack the expertise to interpret or document hardware-level parameters essential for reproducibility.

Unlike many classical workflows, where underlying infrastructure variability often affects performance rather than correctness, in quantum systems hardware-level details may directly alter statistical conclusions. This increases the need for cross-disciplinary collaboration among physicists, HPC specialists, and domain experts.

Quantum computing remains a relatively niche and resource-constrained field [27]. Access to advanced hardware is limited, often concentrated within select research institutions and private companies. Documentation of complete workflows, including calibration data and low-level execution metadata, is frequently incomplete [28]. As in other limited-access fields such as particle physics, verification often depends on scarce resource allocation. Access time is typically prioritized for novel experimentation rather than replication

Moreover, the high-profile status of quantum as a potentially transformative technology and quantum advantage claims create pressure to publish breakthrough results, prioritizing novelty over methodological rigor needed for reproducibility. Such dynamics can create environments where practical incentives work against transparent research despite theoretical commitment to reproducibility [28]. This reflects a larger systemic issue where scientific discourse prioritizes revolutionary potential without adequately addressing result limitations. While expectations management may improve as the field matures, establishing a research culture that prioritizes reproducibility remains essential.

### 3.3 The Meaning of Reproducibility in Quantum Computing

Altogether, having examined the technical, human, and organizational characteristics of quantum computing, we now address one of the questions raised: *How does reproducibility transforms with quantum?*

Where classical reproducibility emphasizes logical conservation of deterministic workflows, quantum reproducibility must contend with the abovementioned problems: statistical rather than deterministic outputs, hardware-dependent physical variability, device drift over time, immature standardization, limited accessibility, etc.

As the quantum computing field matures, developing new foundations of reproducibility that recognizes quantum-specific characteristics, while maintaining connection to the broader computational context becomes timely. *Table 2* presents a summary of the examined quantum characteristics and their implications for reproducibility efforts.

*Table 2.* Quantum Computing Characteristics and Their Implications for Reproducibility

| Quantum Characteristics | | Implication for Reproducibility | Nature of Challenge |
|---|---|---|---|
| Low Technology-Readiness Level (TRL) | **NISQ Era**<br>• TRL 3: Proof-of-concept to prototype transition<br>• 50-100+ noisy qubits without error correction<br>• Operational procedures in flux<br>• Lack of standardization and agreed-upon benchmarks | • Results highly sensitive to specific hardware implementations<br>• Limited cross-device reproducibility<br>• Metadata often missing<br>• Rapidly evolving domain complicates standardization | **Temporary:** Expected to diminish as technology matures |
| Hardware | **Variability**<br>• Multiple competing architectures (superconducting, trapped ions, photonics, etc.)<br>• Device-specific noise profiles<br>**Noise Sources**<br>• Decoherence (T1, T2)<br>• Gate errors<br>• Readout errors<br>• Crosstalk | • Results remain device-dependent, not directly comparable across hardware types<br>• Same circuit yields different outcomes on different devices or over time<br>• Environmental sensitivity affects outcomes | **Mixed:** Architecture diversity may persist for different applications. Noise expected to improve but probabilistic nature requires new frameworks |
| Software | **Software-Hardware Dependency**<br>• Code compiles differently per hardware architecture<br>• Transpilers optimize for specific devices<br>• Version dependencies add complexity | • Cross-platform reproducibility limited<br>• Same high-level code produces different implementations<br>• Comprehensive metadata required for entire software stack | **Persistent:** Inherent to quantum computing's architecture-specific optimization needs |
| Human & Cultural Factors | **Blackbox Problem**<br>• Interdisciplinary field creates knowledge gaps<br>• Domain experts struggle with hardware/software documentation<br>**Niche Field**<br>• Limited expertise and hardware availability<br>• Hardware concentrated in few institutions<br>**Incentive Misalignment**<br>• Pressure for breakthrough results prioritizes novelty over documentation | • Insufficient documentation of workflows and configurations<br>• Cross-disciplinary knowledge gaps hinder comprehensive documentation<br>• Proprietary toolchains limit openness<br>• Verification bottlenecked by hardware access<br>• Novel research prioritized over reproduction | **Mixed:** Expertise and access issues temporary (improving with field growth). Interdisciplinary complexity and incentive structures more persistent |

The last column distinguishes between temporary challenges that will likely resolve as quantum computing matures and those that are mixed or persistent, inherent to quantum computing itself. This temporal distinction matters in the context of developing targeted responses for validating quantum-powered research. While the temporal distinction helps identify which challenges may resolve with technological maturity, quantum computing's persisting characteristics will demand reconceptualizing how reproducibility is defined and practiced.

### 3.4 Revisiting Classical Reproducibility

*Table 3* compares how each reproducibility dimension, including robustness, axes, and infrastructure variation, can be conceptualized across classical and quantum computing respectively. Firstly, quantum computing challenges both aspects of robustness. Classical PR assumes procedures can be standardized and stabilized, while quantum's current state makes this untenable. Hardware-specific noise profiles evolve over time, gate implementations vary across devices, and compilation stacks transform identical high-level code into different physical operations. The consequence is that while high-level scientific questions remain the same, actual quantum code cannot be directly reproduced. As PR serves as a precondition for RR, both must be reconsidered to account for hardware variability. Moving forward, PR should shift from procedural stability toward bounded variability with specific benchmarks across hardware architectures, with RR being bounded within confidence intervals.

When it comes to the axes of reproducibility, *transparency* should be extended to include detailed hardware characterization, calibration parameters, and compilation configurations, as at the current maturity stage, results cannot be separated from the specific hardware that produced them. *Portability* likewise faces fundamental limitations: since identical execution is infeasible with quantum, developing standardized benchmarks for confidence intervals that account for hardware variability becomes an imperative. *Coverage* in quantum must encompass both software and hardware elements (hardware descriptions, error mitigation techniques, circuit optimization) as they are mutually dependent at this maturity level. Lastly, infrastructure variation becomes blurred, as hardware and noise profiles directly impact research results.

*Table 3.* Reproducibility Across Classical and Quantum Computing, with Proposed Reconceptualization

| Dimension | Classical Computing | Quantum Computing | Proposed Reconceptualization |
|---|---|---|---|
| **Robustness** | | | |
| Procedure Robustness (PR) | • Procedures can be standardized<br>• PR defined as [...] | • Limited by hardware-specific noise profiles | PR as bounded variability of procedure |
| Results Robustness (RR) | • Deterministic computation produces consistent results<br>• Differences in results indicate errors | • Inherently probabilistic nature means results are distributions | RR as bounded within confidence intervals, needs to be integrated with PR |
| **Axes of Reproducibility** | | | |
| Transparency | • Documentation needs focus primarily on software aspects as hardware is mature | • Requires hardware characterization & noise characteristics due to low TRL | Inclusion of detailed hardware characterization, calibration parameters, and compilation configuratins |
| Portability | • Code can generally run consistently across different systems, with containerization enabling portability | • Limited by hardware-specificity, device-specific noise profiles & execution | Develop benchmarking of statistical consistency, as identical execution not possible |
| Coverage | • Workflows with relatively clear input-output relationships<br>• Intermediate computation states can be captured & verified<br>• The focus is on sofrware pipeline documentation | • Requires documentation of error mitigation techniques & hardware calibration | Focus on both software & hardware |
| **Infrastructure Variation** (Same setup condition) | • Mature software & hardware: Configurations and versions can generally be well-documented and reproduced<br>• Generally, a relatively clear distinction between infrastructure variation and reproducibility | • Low TRL: Device-specificity & hardware has a direct impact on scientific results<br>• Infrastructure variations then become an integral part of scientific results<br>• By now, collapsed distinction between infrastructure variation and reproducibility | Move toward Meta-workflows |

Moving forward, upon classical and quantum computing integration, classical deterministic processes intersect with quantum probabilistic behavior. Beyond the distinct challenges from each paradigm, their integration may amplify existing issues or generate entirely new ones.

## 4. Quantum as a Classical Computing Accelerator

The future of scientific computing increasingly points toward hybrid systems where quantum computers serve as specialized accelerators within broader HPC infrastructures. Their integration represents the next era of computational science [1], promising to solve ever-complex scientific problems, beyond the reach of either approach alone. The vision of hybrid computing is to create workflows where scientific problems are decomposed according to each computation type's strengths [1], smoothly transitioning between each other. Understanding such transitions is detrimental to developing what reproducibility means in the emerging context.

### 4.1. New Reproducibility Challenges at the Classical-Quantum Integration

While quantum computing's TRL timeline suggests gradual evolution toward fault-tolerant systems, reaching high maturity levels may take around 10 years [12]. Higher maturity will resolve some challenges, while others remain inherent to quantum computation. During this transitional period, several key aspects require particular attention:

**Probabilistic-Deterministic Integration:** Quantum components produce probability distributions that vary between runs, while classical components yield deterministic results. Hybrid workflows must accommodate both paradigms simultaneously, requiring new frameworks for defining procedure robustness that can handle fundamentally different output types.

**Hardware Abstraction Gaps:** Classical components remain portable through mature abstraction layers, while quantum components stay hardware-bound, with tight connection between high-level quantum code and specific hardware architectures creating barriers to the portability dimension of reproducibility.

**Knowledge Silos:** The knowledge gaps between classical and quantum computing experts create challenges for developing reliable documentation patterns. This reflects a broader issue of fragmentation in computational science [10], where increasing complexity distributes expertise across multiple domains. During the upcoming period, raising awareness about these cross-paradigm gaps thereby becomes a key action for developing reliable documentation patterns.

**Verification Limits:** Reproducibility concerns intensify when quantum advantage is claimed and classical verification becomes computationally infeasible. This is already evident in quantum advantage claims by major technology companies [28], which remain difficult to validate due to limited hardware access and computational constraints. While resource-intensive validation is not unique to quantum (experimental fields like particle physics face similar limitations), hybrid computing with quantum components operating beyond classical computational reach create verification bottlenecks, as independent reproduction requires access to similarly capable hardware.

Altogether, the classical-quantum integration creates qualitatively new complications that cannot be resolved through paradigm-specific approaches alone. The outlined reproducibility challenges across the previous sections may be symptomatic of a broader issue: the increasing complexity of computational research is outpacing our methodologies for ensuring reproducibility. In this light, rather than treating it as an isolated challenge, quantum-classical integration can be approached as a catalyst exposing deeper issues about how computational research is structured and validated.

### 4.2. Quantum Integration as Symptom of Broader Issue in Computation

Examining reproducibility challenges across classical and quantum computing reveals a telling pattern. Several reproducibility limitations presented as quantum-specific are not fundamentally new, but seem to reflect long-standing tensions in computational sciences that have intensified with growing complexity. Hardware-software dependencies illustrate this clearly. While a prominent issue in quantum computing, these dependencies have long existed in classical computing environments but were often overlooked [9] [10]. With increasing computational complexity, hardware independence becomes harder to maintain, with documentation practices struggling to evolve accordingly. Quantum computing, with its unavoidable hardware dependencies, makes explicit what affects all computational science - results can become inseparable from the specific computational environments producing them. When results cannot be reliably reproduced across different computational environments, the scientific community loses its primary mechanism for validating claims.

Documentation challenges for quantum workflows also expose broader inadequacies in how computational research is recorded and communicated. As computational methods become more sophisticated, the knowledge required to understand and reproduce experiments exceeds what traditional scientific publications can reasonably capture within incentive structures that already limit comprehensive documentation. This increasing specialization needed across infrastructure components – from algorithm design to optimization – distributes reproducibility expertise across multiple domains.

In this way, some of the approaches being developed for examining quantum reproducibility may ultimately benefit all computational domains. Despite quantum computing's unique characteristics, the quantum-classical integration could serve as a catalyst for deeper methodological examination. The path forward involves transforming how computational science itself is structured and validated, which begins by recognizing workflows as foundational to scientific method rather than technical afterthoughts. Such a shift could then open pathways for addressing the reproducibility challenges identified across classical, quantum, and hybrid computing paradigms.

## 5. Workflow-Centered Science: A Path Forward

The reproducibility challenges examined across classical, quantum, and hybrid computing converge on a central issue: computational complexity has outpaced the methodologies available to capture and validate it. With experiment results increasingly inseparable from the computational environments producing them, maintaining reproducibility demands systematic approaches for abstracting research processes from infrastructure implementation details. Workflow-centered science provides this systematic approach, positioning workflows as the primary mechanism for separating what must remain constant for validity from what can vary across implementations.

Turning scientific ideas into executable workflows requires decomposing research problems into discrete computational tasks, identifying dependencies, and formalizing them in machine-executable form. This translation requires both domain knowledge and workflow engineering skills, as researchers alone often struggle to bridge this gap [10]. One of the main implications of this workflow co-creation is a change in roles across the scientific ecosystem. Domain researchers evolve from being passive tool users to active workflow co-designers who, together with the infrastructure providers, shape how computational resources support their research. This collaboration requires new skills, an open mindset, and practices that go beyond traditional disciplinary boundaries. A long-term effort to develop a cross-disciplinary literacy to collaborate could be a welcome first step to acknowledge that modern computational science exceeds the expertise of any individual researcher, requiring collective knowledge for robust reproducibility.

## 5.1. Meta-Workflows: Abstracting Research Logic from Infrastructure

At the core of workflow-centered science are *meta-workflows*, a collection of sub-workflows containing multiple tasks that may span across multiple domains and include workflows from heterogeneous workflow platforms [9] Rather than encoding workflows for particular systems, such as specific HPC clusters or a particular quantum device architecture, meta-workflows describe what the research aims to accomplish, and the dependencies between computational steps, and the validity requirements that implementations must satisfy to do so.

As quantum and classical infrastructures evolve, with ever-changing hardware architectures, software stacks updates, and access mechanisms shifts among others, meta-workflows abstract the scientific questions and methodological approaches from particular hardware or software. This enables research to adapt without requiring complete redesign.

For instance, a meta-workflow might specify that quantum circuit depth must remain below a threshold for noise considerations, while leaving the specific quantum processor architecture as an implementation detail that can vary. While such specification is subject to the relatively low TRL of quantum computing, as quantum matures toward fault-tolerant systems, meta-workflows can evolve to capture increasingly stable validity requirements while maintaining abstraction from specific hardware implementations. When workflows document which infrastructure aspects matter for scientific validity versus which constitute implementation details, they preserve expertise that would otherwise remain tacit or could be lost when translated to different computing environments.

## 5.2. What Happens to Scientific Knowledge?

As quantum and other advanced classical computational approaches, such as HPC, transform how science is conducted, we must also reconsider what constitutes valid scientific knowledge – and how we establish confidence in scientific claims. By now, the assumption has been that knowledge claims must be testable and verifiable by others, with reproducibility being one of the cornerstones of scientific validity. Workflow-centered science does not abandon this principle, but it brings change into how it is realized, especially in the context of the increasingly complex computational environments. The nature of scientific knowledge itself begins to shift into a more contextual understanding of the process.

Meta-workflows abstract scientific processes away from specific implementations and create a higher-level representation of research that can survive technological changes. However, they simultaneously also make explicit how results depend on the specific computational environments that produced them. The abstraction happens at the level of the scientific process, while the contextualization happens at the level of interpreting specific results. A well-designed meta-workflow can be implemented across different quantum or classical systems, but the results from each implementation are to be interpreted within their specific computational context.

In this light, this shift can be understood as recognizing that knowledge emerges from structured interactions between researchers, tools and methodologies. Scientific knowledge depends both on the questions we ask and the means through which we investigate them. By developing meta-workflows that document both process abstractions and implementation contexts, we create a more robust foundation for scientific knowledge that acknowledges complexity without sacrificing rigor. The path forward involves embracing this evolution in understanding scientific knowledge rather than resisting it.

## 6. Conclusion

This paper examined reproducibility challenges across classical computing, quantum computing, and their integration. While classical and quantum computing each face distinct reproducibility challenges, from hardware dependencies and documentation inadequacies in classical computation to probabilistic behavior and tight software-hardware coupling in quantum computation, their integration exposes a pattern, where computational complexity outpaces the methodologies available to ensure scientific rigor.

The assumption that computational infrastructure serves as a neutral, reproducible platform seems to collapse, as results can become inseparable from the specific computational environments producing them. This inseparability demands new approaches that can abstract research logic from implementation details while explicitly capturing dependencies that matter for validity. Workflow-centered science provides this, with meta-workflows enabling research to remain stable across evolving computational environments. This requires reconceptualizing roles: researchers evolving from passive tool users to active co-designers collaborating with infrastructure providers, acknowledging that advanced computational science exceeds the expertise of individual domains. This helps position the scientific community to develop solutions that strengthen reproducibility across all computational domains.

As computational science continues to advance in complexity, with quantum computing representing just one frontier of this advancement, our conceptions of reproducibility, validation, and knowledge must advance as well. By recognizing quantum integration as a catalyst for this epistemic evolution, we position ourselves to develop not just better technical solutions for reproducibility but more sophisticated frameworks for understanding what constitutes valid scientific knowledge in an era of increasingly complex computational science. In this way, the reproducibility challenges presented by quantum-HPC integration become an opportunity for strengthening rather than foundations of scientific knowledge.